\documentclass[twocolumn,amsmath,amssymb,prl,superscriptaddress,showpacs]{revtex4}
\usepackage{graphicx}
\usepackage{bm}

\begin{document}

\title{Defect Formation Preempts Dynamical Symmetry Breaking in Closed 
Quantum Systems}

\author{Carmine  Ortix}
\affiliation{Institute for Theoretical Solid State Physics, IFW-Dresden, D01171 Dresden, Germany}
\author{Jorrit  Rijnbeek}
\affiliation{Insitute-Lorentz for Theoretical Physics, Universiteit Leiden, 2300 RA Leiden, The Netherlands. }
\author{Jeroen van den Brink} 
\affiliation{Institute for Theoretical Solid State Physics, IFW-Dresden, D01171 Dresden, Germany}

\date{\today}

\begin{abstract}
 We show that no matter how slowly a quantum-to-classical symmetry breaking process is driven, the adiabatic limit can never be reached in a macroscopic body. Massive defect formation preempts an adiabatic quantum-classical crossover and triggers the appearance of a symmetric non-equilibrium state that recursively collapses into the classical state, breaking the symmetry at punctured times. The presence of this state allows the quantum-classical transition to be investigated and controlled in mesoscopic devices by supplying externally the proper dynamical symmetry breaking perturbation.
\end{abstract}

%%03.65.-w Quantum mechanics
%% 11.30.Qc Spontaneous and radiative symmetry breaking
%% 64.60.Ht Dynamic critical phenomena
%%  75.10.-b General theory and models of magnetic order 

\maketitle

{\it Introduction}. -- The relation between quantum physics at microscopic scales and the classical behavior of macroscopic bodies has been debated since the inception of quantum theory. The fundamental difference is that while in quantum mechanics all configurations equivalent by symmetry have the same status, in classical physics one of them is singled out -- the symmetry is broken.
This spontaneous symmetry breaking causes a macroscopic body under equilibrium conditions to have less symmetry than its microscopic building blocks~\cite{and72}. Superconductors, antiferromagnets, liquid crystals, Bose-Einstein condensates and crystals all exhibit spontaneously broken symmetries. The general idea is that when the number of microscopic quantum constituents, which depending on the system corresponds to the number of Cooper pairs, particles or spins, goes to infinity the quantum  system undergoes a phase transition into a state that violates the microscopic symmetries. From a purely theoretical perspective, spontaneous symmetry breaking is thus related to a singularity of the thermodynamic limit. If this thermodynamic singularity is present, a symmetry broken groundstate exists. This observation can be formalized into an exact statement on the {\it existence} of a broken symmetry state~\cite{and52,wez05}, but it makes no assertion on whether or how it can evolve out of  the symmetric state, nor on the dynamics of quantum-to-classical transitions. So the question remains how a continuous symmetry can be broken {\it dynamically}. 

We investigate this by considering a symmetry breaking field that slowly drives an arbitrary large but finite system from quantum to classical. If the symmetry breaking process were fully adiabatic, the effect of the driving would correspond to subjecting the system to a quasi-static symmetry breaking field. In this case the timescale at which the classical state is singled out and symmetry is broken becomes shorter and shorter as the size of the system grows. However, that the time-evolution be adiabatic is not evident. The adiabatic theorem states that under slow enough external perturbations, there are no transitions between different energy levels. When the distance between energy levels is exponentially small, the adiabatic evolution is hampered and transitions between levels become unavoidable. Such transitions and their associated defects can in principle strongly affect the symmetry breaking process. Here we show that no matter how slowly a symmetry breaking perturbation is driven, the adiabatic limit cannot be reached. Defect formation turns out to be so pervasive that it preempts an adiabatic quantum-classical crossover in macroscopic systems. The existence of this non-adiabatic regime is consistent with the recent discovery~\cite{pol08} that adiabatic processes in low dimensional systems with broken continuous symmetries are absent.  

The far-from-equilibrium time-evolution caused by a symmetry breaking field has remarkable consequences. We will show that in any large finite system the non-equilibrium state does not break the symmetry. However, it recursively collapses into the purely classical state: it breaks the symmetry at punctured times, resulting in a Dirac comb of symmetry broken, classical states.  This Dirac comb of quantum-classical transitions can be investigated in mesoscopic devices by supplying a proper dynamical symmetry breaking perturbation. 

%In ultra-cold atom systems, for instance, the necessary experimental prerequisites such as isolation from the environment together with the possibility of real-time control of system parameters are fulfilled. 

{\it Spontaneous symmetry breaking}. -- Even if {\it a priori} spontaneous symmetry breaking  is an intractable problem involving a near infinity of interacting quantum degrees of freedom, there is a representative, integrable model that exhibits spontaneous symmetry breaking: the Lieb-Mattis model~\cite{lie62}. It is the effective collective Hamiltonian that underlies the breaking of the SU$(2)$ spin rotation symmetry in generic Heisenberg models with short-range interactions. Very similar collective models underlie the breaking of other continuous symmetries as the gauge invariance in superconductors or the translational symmetry in quantum crystals~\cite{wez08}. The results that we will present here are therefore robust and generic.

The quantum-classical symmetry-breaking transition is manifest in the Lieb-Mattis model once a symmetry breaking field $H$, in this case a staggered magnetic field, is introduced. Before turning to dynamical symmetry breaking we first summarize a few essential features of the Lieb-Mattis Hamiltonian. The Hamiltonian is defined for spins $1/2$ on a bipartite lattice with sublattices $A$ and $B$, where ${\bf S}_{A,B}$ is the total spin on the $A/B$ sublattice with $z$ projection $S_{A/B}^z$: 
\begin{equation}
{\cal H}_{LM}=\dfrac{2 \left| J \right|}{N} {\bf S}_A \cdot {\bf S}_B- H \left(S_A^z-S_B^z \right).
\label{eq:liebmattishamiltonian}
\end{equation}
 Every spin on sublattice A interacts with all spins on sublattice B and vice versa with an interaction strength $2 \left|J\right| / N$ (which depends upon the number of sites $N$). Taking $H=0$, the model can be solved by introducing the total spin operator ${\bf S}={\bf S}_A+{\bf S}_B$. The eigenstates of the Hamiltonian  are then $|S_{A},S_B,S,M \rangle$ where $S$, $M$ indicate the total spin and its $z$ axis projection, whereas $S_{A,B}$ are the total sublattice spin quantum numbers. The ground state is symmetric and corresponds to an overall $S=0$ singlet with $S_{A,B}$ maximally polarized and is characterized by zero staggered magnetization. Magnon excitations carry an energy $J$ and are realized lowering either $S_{A}$ or $S_{B}$. The $S \neq 0$ quantum numbers label a tower of states  with energy scale $E_{thin}=J / N$, which is also referred to as the thin spectrum. It is {\it thin} because it contains states that are so sparse and of such low energy that their contribution to thermodynamic quantities vanish in the thermodynamic limit~\cite{wez08}. Nevertheless, when $N \rightarrow \infty$ the thin spectrum excitations collapse and form a degenerate continuum of states. Within this continuum, even an infinitesimally small symmetry breaking perturbation $H$ is enough to stabilize the fully ordered symmetry broken ground state -- the system is inferred to spontaneously break its symmetry. The finite symmetry breaking field $H$ couples the thin spectrum states so that the eigenstates $|n \rangle= \sum_S u^n_S |S \rangle$ of the Lieb-Mattis model become wave packets of total spin states. In the continuum limit, where $N$ is large and $0 \ll S \ll N$, the corresponding low-energy effective Hamiltonian is~\cite{wez05}
\begin{equation}
{\cal H}=\dfrac{H \,N}{4\, \hbar^2} \,\Pi^2+\dfrac{J}{N} S^2,
\label{eq:continuumhamiltonian}
\end{equation}
where $\Pi$ is the conjugate momentum of the total spin $S$. The eigenstates $u^n_S$ are harmonic oscillator states of order $n$, with $n$ odd in order to meet the boundary condition $S \geq 0$. One can easily show the singular nature of the thermodynamic limit in the $n=1$ ground state by calculating the expectation value of the order parameter  $2 \,\langle S_A^z-S_B^z \rangle \sim N \, \mathrm{e}^{- \omega_S}$ where the dimensionless parameter $\omega_S= N^{-1} \sqrt{4 J / H }$.  
When sending first $H \rightarrow 0$ and then $N \rightarrow \infty$, the singlet state appears as the ground state, which respects the spin rotational symmetry,  {\it i.e.} $2 \langle S_A^z-S_B^z \rangle \equiv 0$. Taking the limits in opposite order, one finds that the ground state corresponds to the fully polarized antiferromagnetic Ne\'el state with a fully developed order parameter $2 \langle S_A^z-S_B^z \rangle \equiv N$. In this case $E_{thin}=\sqrt{J \, H }$ represents the typical energy of the excitations labeled by $n$ that now act as a {\it dual} thin spectrum.

\begin{figure}
\includegraphics[width=\columnwidth]{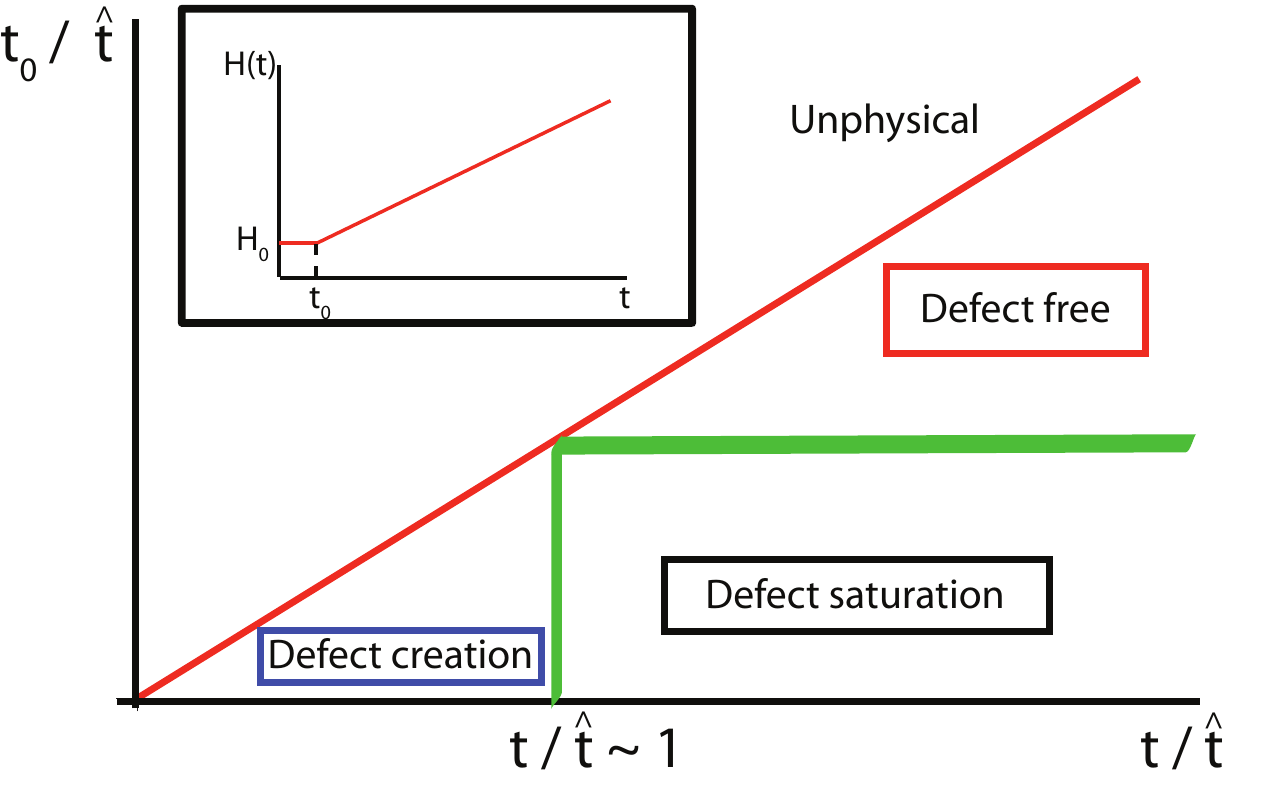}
\caption{ The three different regimes for the behavior of the density of defects in the $t-t_0$ plane. Times have been measured in unit of the freeze out time $\hat{t}$. The bold lines indicate the crossover among the different regimes whereas the straight line limits the physical region with $t>t_0$. The inset shows the setup of the symmetry breaking field considered throughout this work.}
\label{fig:phasediagram}
\end{figure}

\begin{figure}
\includegraphics[width=\columnwidth]{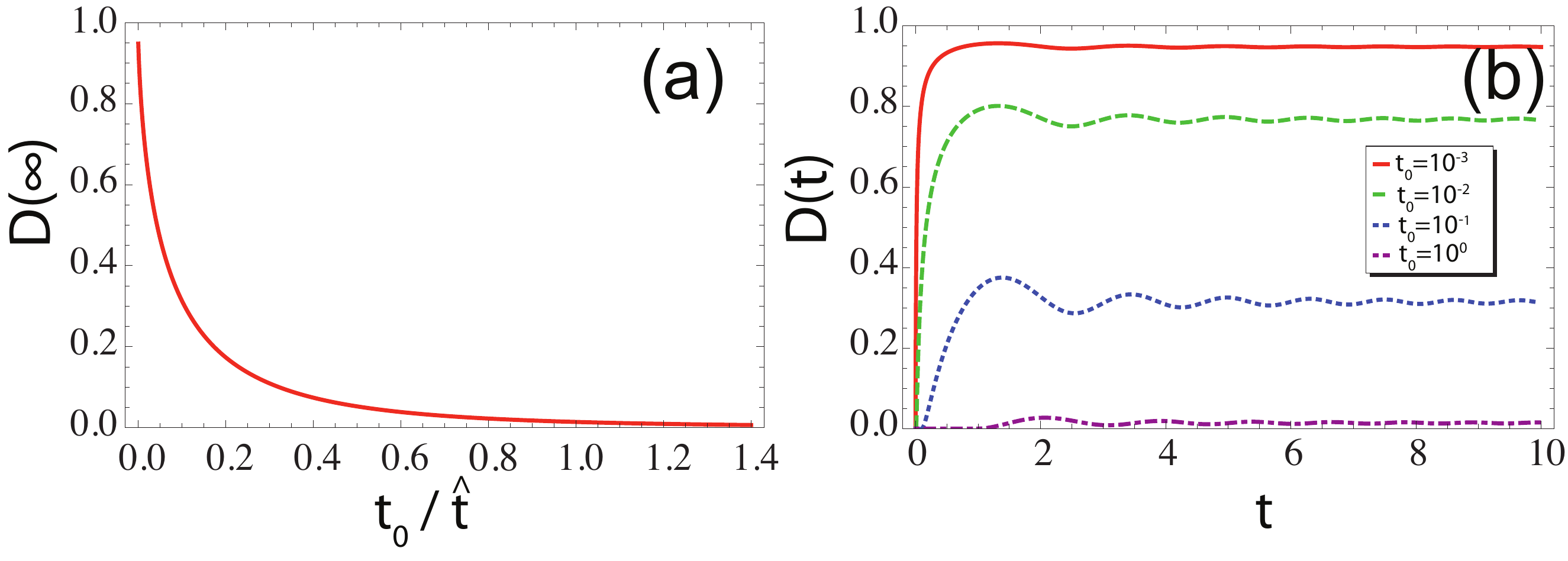}
\caption{(a) Asymptotic value of the density of defects as a function of the initial time $t_0$ over freeze-out time ${\hat t}$. (b) Time evolution of the density of defects for different values of the initial time $t_0$. Times are measured in units of the freeze-out time ${\hat t}$. The curves are independent of $N$.}
\label{fig:defects}
\end{figure}

{\it Adiabatic-impulse approach}. -- Let us now consider the dynamical case and turn on the symmetry breaking field linearly in time $H(t)=\delta \, t, $
with ramp rate  $\delta$. At initial time $t_0$ we start out with a field $H(t_0)=H_0$ [see the inset of Fig.~\ref{fig:phasediagram}] and the wavefunction of the system corresponding to this static ground state. We introduce $H_0$ in order to have a cutoff that guarantees the continuity of the wavefunction basis. Lateron we will consider the limit $H_0 \rightarrow 0$ which corresponds to an initial state that is a completely symmetric singlet.
To capture the dynamics of the symmetry-breaking transition we first use the quantum Kibble-Zurek (KZ) theory~\cite{zur05,pol05}. The essence of the KZ theory of non-equilibrium phase transitions~\cite{kib80} is a splitting of the dynamics into a nearly critical impulse regime where the system's state is effectively frozen and a quasi-adiabatic regime far from the critical point where transitions among the instantaneous eigenstates of the Hamiltonian are neglected. This splitting defines the so-called adiabatic-impulse approximation~\cite{dam06}. In particular, the critical impulse regime occurs whenever the characteristic relaxation time $\tau(t)=\hbar / E_{thin}(t)$  is much larger than the timescale  $t$ on which the Hamiltonian is changed. On the contrary for $\tau(t) << t$, the system's state is able to adjust to the changing symmetry breaking field, and the transitions among the dual thin spectrum excitations can be neglected. The crossover between the two regimes is determined by Zurek's equation~\cite{zur05} $\tau(\hat{t})=\hat{t}$ and defines the freeze-out time 
\begin{equation} 
\hat{t}= \left[ \frac{\hbar^2 }{J \, \delta} \right]^{1/3}. 
\end{equation}
For an initial time $t_0 \gg \hat{t}$, the system's dynamics will thus be nearly adiabatic [c.f. Fig.~\ref{fig:phasediagram}]. 
Strictly speaking, in the true adiabatic limit ($t_0 / \hat{t} \rightarrow \infty$) the probability of switching thin spectrum levels will be vanishingly small. To quantify this, we calculate the fidelity of the snapshot ground-state wavefunction~\cite{dor03} ${\it f}(t)=|\,\langle \,u_S^1(t) \,\,|\,\, \psi(t) \,\rangle \,|^2$ with $\psi(t)$ the actual ground-state wavefunction and the associated density of defects  ${\cal D}(t)=1-{\it f}(t)$ [c.f. Fig.~\ref{fig:defects}]. The ramp rate of a nearly defect-free quench is seen to be bounded by $\delta \ll \sqrt{ H_0^3 \, J} / \hbar .$ Therefore the limits $\delta \rightarrow 0$ and $H_0 \rightarrow 0$ do not commute. In other words: no matter how slowly one drives the symmetry breaking field, if the initial symmetry breaking field $H_0$ is sufficiently small, the adiabatic limit can never be reached. 

Besides just an adiabatic time-evolution, the KZ analysis renders two non-trivial regimes for the dynamics of an initially symmetric ground state [c.f. Fig.~\ref{fig:phasediagram}]. First the evolution takes place in the impulse regime ($ t_0 \ll  t \ll  \hat{t} \,$)  where the spin singlet state is effectively frozen and changes only by a trivial overall phase factor. In this regime, the density of dynamically generated defects grows continuously in time [c.f. Fig.~\ref{fig:defects}(b)]. This evolution lasts until $t>\hat{t}$. At the freeze-out time $\hat{t}$, defect formation stops and the defect density saturates [c.f. Fig.~\ref{fig:defects}] to ${\cal D}(\hat{t}) \sim 1- 8 \left( t_0 / \hat{t} \,\, \right)^{3/4}$ . The defect density thus tends arbitrarily close to $1$  for small enough initial symmetry breaking field $H_0$. In this case the system actually reaches a state that is a superposition of an exponentially large number of thin spectrum excitations. The quantum-classical crossover is thus accompanied by massive defect formation, which in the end preempts such a crossover. In the subsequent adiabatic regime $t \gg \hat{t}$ no new excitations are created. The corresponding evolution of the wavefunction can be analyzed as follows:
we first expand the frozen initial ground state as a superposition of the dual thin spectrum eigenstates at the freeze out time ${\hat t}$ as
$
u_S^1(t_0)=\sum_n c_n u_S^n (\hat{t} \,),
$
where the coefficients $c_n$ are non zero only for odd values of the quantum number $n$. Since the evolution is considered to be adiabatic, 
we may write the time evolution of the wavefunction as
$\psi_S(t)=\sum_n c_n \,u_S^n(t) \, \mathrm{e}^{i\, \gamma_n(t)} \, \mathrm{e}^{-i \Omega_n(t)},$
where $\gamma_n(t)$ is the Berry phase~\cite{ber84} associated to the eigenstate $|n \rangle$ which turns out to be zero and we have defined the dynamical phase
$
\Omega_n(t)=\frac{2}{3}  \left(n+ \frac{1}{2} \right) \left[\left( t / \hat{t} \, \right)^{3/2}-1 \right].
$
As the time increases, the various thin spectrum eigenstates all pick up a different dynamical phase leading to quantum interference. However for  $t_k=(1+ \frac{3 k \pi}{2} )^{2/3} \, \hat{t}$ with $k$ integer, the interference is fully constructive and the wavefunction corresponds to the instantaneous ground state $u_S^1(t \, t_0 / \, \hat{t})$ as easily follows by considering that the coefficients $c_n$ depends on the ratio $t_0 / \hat{t}$ alone. The system's state then corresponds precisely to the snapshot ground state of a symmetry broken Lieb-Mattis model subject to a {\it renormalized} staggered magnetic field $H_R= H \times H_0 / [\hbar^2 \, \delta^2 / (2 \, J  )]^{1/3}$. Thus when the initial symmetry breaking field $H_0$ vanishes, the symmetric singlet state becomes fact at any recursion time. 

\begin{figure}
\center{\includegraphics[width=\columnwidth]{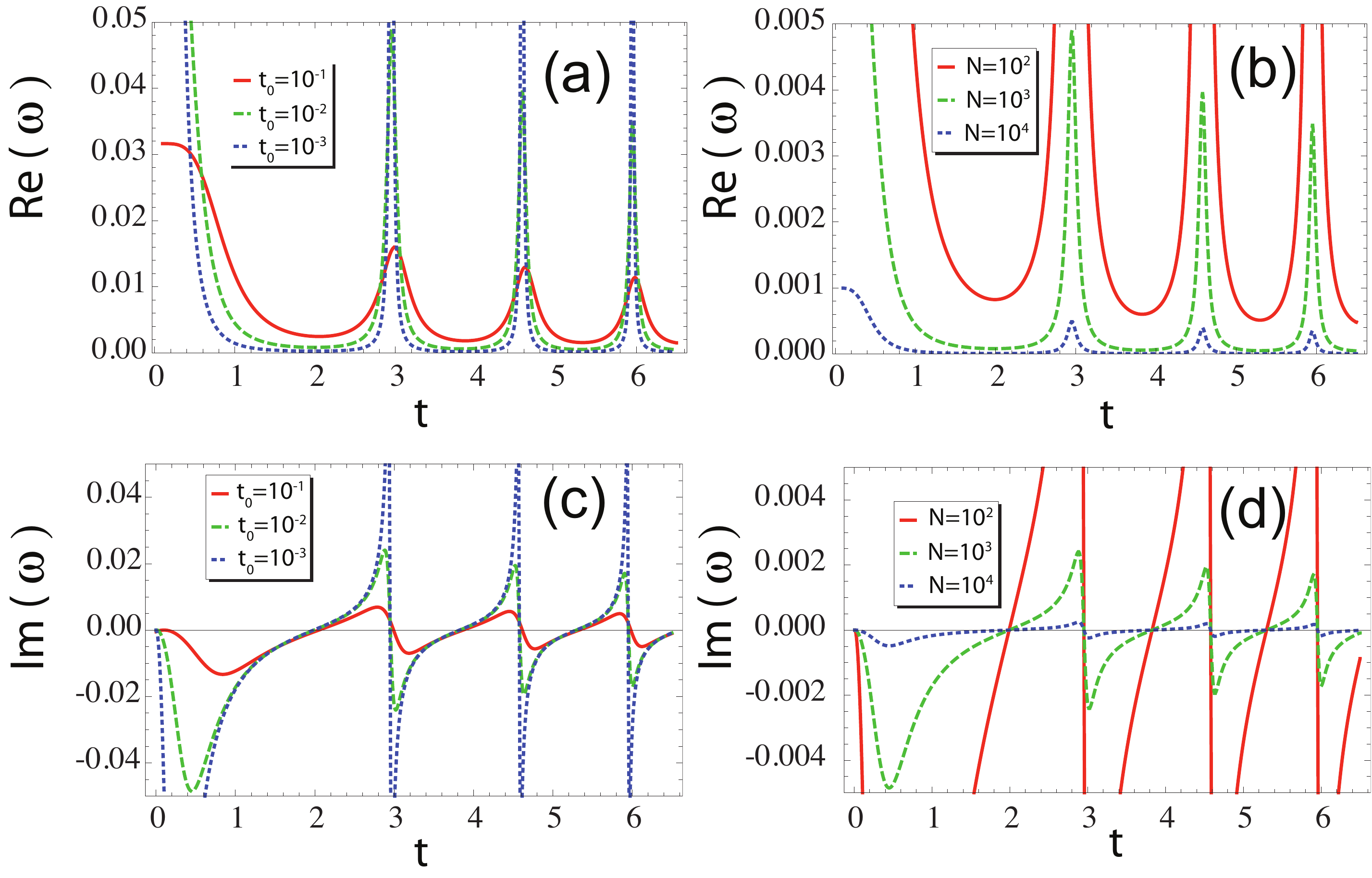}}
\caption{(a) Time-dependence of the real part of the dimensionless parameter $\omega$ for different values of the initial time $t_0$ for $N=10^2$ spins and a freeze-out time $\hat{t}=1$. All times are measured in units of $4 J / \delta$. (b) Same for a fixed initial time $t_0=10^{-2}$ and different values of the number of sites $N$. (c),(d) Same as panels (a) ,(b) for the imaginary part of the dimensionless parameter $\omega$.}
\label{fig:realpart}
\end{figure}

\begin{figure}
\center{\includegraphics[width=\columnwidth]{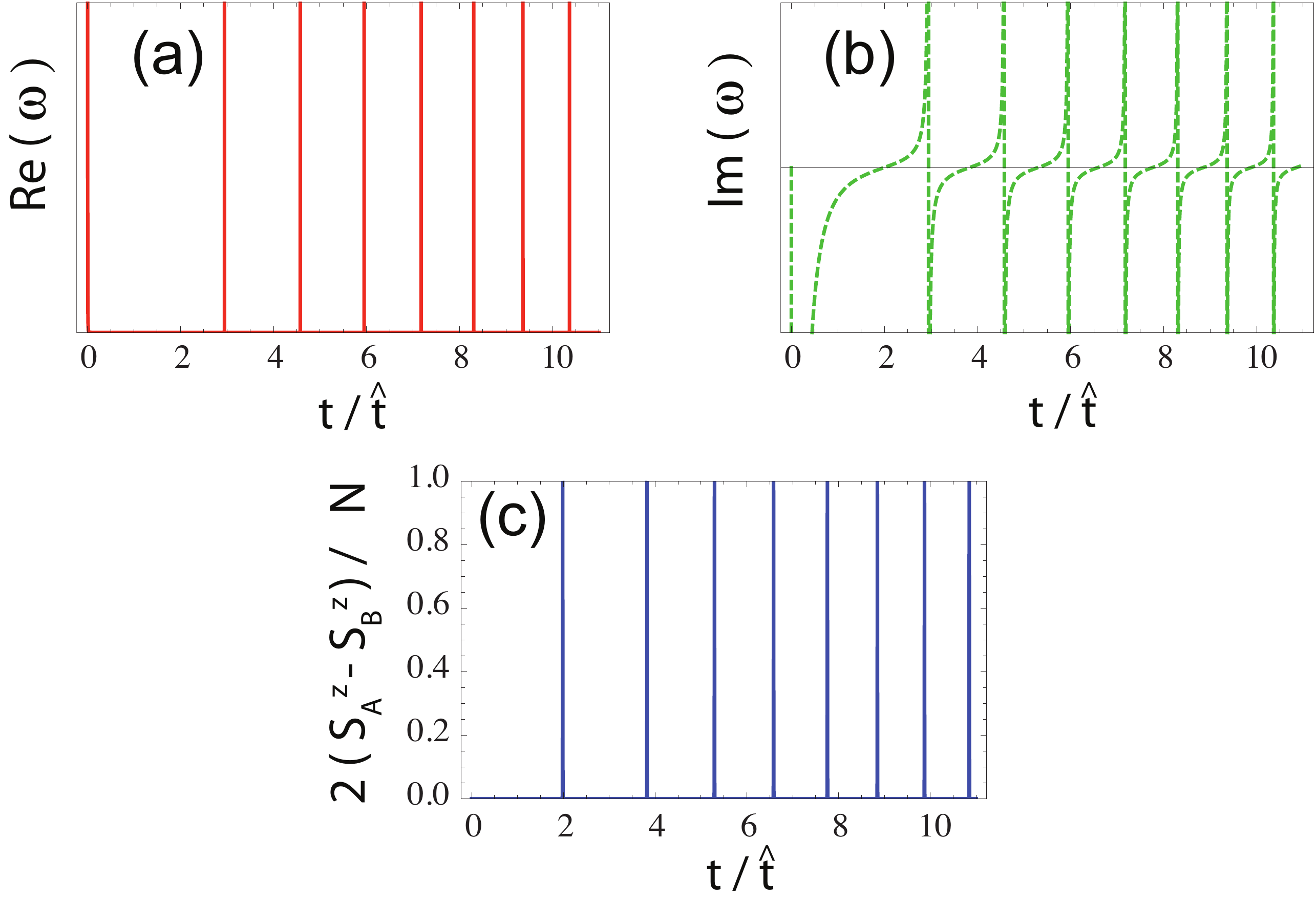}}
\caption{Asymptotic behavior of the real (a) and the imaginary (b) part of the dimensionless parameter $\omega$ in the limit $t_0 \rightarrow 0$. (c) Asymptotic behavior of the order parameter as a function of time measured in units of the freeze-out time $\hat{t}$.}
\label{fig:analytical}
\end{figure}

{\it Exact quantum theory}. -- A full description of the interference effects in a highly non-adiabatic state within the adiabatic-impulse method is in practice impossible. 
We can, however, explicitly monitor quantum phase interference effects by constructing the {\it exact} non-equilibrium wavefunction for Hamiltonian Eq.~(\ref{eq:continuumhamiltonian}). 
Within the Feynman path integral approach, it can be shown~\cite{son99} that the propagator has a spectral decomposition ${\cal G}(S_B, t_B| S_A, t_A)=\sum_n \Psi_{S_A}^{n \, \,\star}(t_A) \,\, \Psi_{S_B}^n(t_B)$ in terms of a complete set of wavefunctions of the form
\begin{eqnarray}
\Psi_{S}^n(t)&=&\sqrt{\dfrac{1}{2^{n-1} \, n!}} \left[\dfrac{{\it Re} \left(\omega (t) \right)}{\pi}\right]^{1/4} \,\, \mathrm{e}^{-\mathrm{i} \left(n+\frac{1}{2}\right) \phi(t)} \, \times \nonumber \\ & &  H_{n}\left[\sqrt{{\it Re}\left(\omega(t)\right)}\, S\right]\,\, \mathrm{e}^{-\frac{S^2}{2} \omega(t)},
\label{eq:dynamicwavefunction}
\end{eqnarray}
where the quantal phase $\phi(t)$ and the complex dimensionless parameter $\omega(t)$ are uniquely determined by the classical Euler-Lagrange equation of motion whereas the quantum number $n$ takes only odd values in order to meet the boundary condition $S \geq 0$. Different sets of wavefunction of the form Eq.~(\ref{eq:dynamicwavefunction}) exist and correspond to take different pairs of linearly independent solutions to the equation of motion. This allows us to choose two particular solutions guaranteeing that at the initial time, $\Psi_S^1(t_0) \equiv u_S^1(t_0)$ so that the exact wavefunction remains an $n=1$ state of the form Eq.~(\ref{eq:dynamicwavefunction}).  
The wavefunction is thus completely characterized by a dimensionless parameter $\omega$ (which plays a similar role as $\omega_S$ in the static case) that now has both a real and an imaginary part. 
The resulting time dependence of the real part of $\omega$ for different values of  $t_0$ is shown in Fig.~\ref{fig:realpart}(a). By decreasing the initial time, it develops a series of sharp peaks 
which eventually leads to a Dirac comb structure in the $t_0 \rightarrow 0$ limit  [c.f. Fig.~\ref{fig:analytical}(a)] with singularities at instants $t_k^R \simeq \hat{t} \left(\frac{3}{2} k \, \pi + \frac{13 \pi}{8} \right)^{\frac{2}{3}}.$
The imaginary part of $\omega$ has a time dependence as in Fig.~\ref{fig:realpart}(c). As the initial time decreases, it approaches a characteristic tangent-like behavior [c.f. Fig.~\ref{fig:analytical} (b)] 
with singularities appearing precisely at the Dirac deltas of ${\it Re}(\omega)$. 
The limiting behavior of $\omega$ is universal since it scales with $N^{-1}$ as the number of sites is varied [c.f. Fig.~\ref{fig:realpart}(b),(d)].   

The exact time development reveals that when the symmetric system is subject to the symmetry breaking field, a non-equilibrium state forms that is intermediate between a pure quantum symmetric and a pure classical state. It is a vast superposition of $ S \neq 0$ thin spectrum excitations, with complex amplitudes. This state {\it does not} break the SU$(2)$  symmetry, as a computation of  $\langle S_A^z-S_B^z \rangle$ directly demonstrates. 
As time evolves, this non-equilibrium state develops smoothly, until at a certain moment the system's state corresponds {\it precisely} to the classical symmetry broken ground state -- the fully polarized  N\'eel ground state of the symmetry broken Lieb-Mattis model. This classical state forms at punctured times 
$
t_k^I \simeq \hat{t} \left(\frac{3}{2} k \, \pi + \frac{7 \pi}{8} \right)^{\frac{2}{3}}
$
 . At any other instant, the spin rotation symmetry is restored. This is in agreement with the 
adiabatic-impulse analysis which does not allow symmetry breaking of a symmetric state when quantum phase interference effects are neglected. 
As a result, the time evolution of the order parameter is characterized by a 
comb structure 
$2 \,\langle S_A^z- S_B^z \rangle = N \,\sum_{k \geq 0} \delta_{t,t_k^I}$ 
[see Fig. \ref{fig:analytical}(c)] which corresponds to the periodic emergence of the symmetry broken states at punctured times. These instants are related to the freeze-out time alone, indicating the non-equilibrium nature of this dynamical symmetry breaking phenomenon.   The freeze-out time can be experimentally tuned by changing the ramp rate of the symmetry breaking field and a quantum-classical transition can be induced in individual mesoscopic quantum objects by supplying a proper dynamical symmetry breaking perturbation. In the case of an infinitely sudden quench ($\delta \rightarrow \infty$), the freeze-out time vanishes and the punctured times of symmetry broken classical states collapse onto each other. In the contrary, asymptotically adiabatic limit ($\delta \rightarrow 0$), the first punctured time of symmetry broken state diverges: the system never breaks its symmetry. 

{\it Conclusions}. -- In the dynamical realm the quantum-classical symmetry breaking transition is thus characterized by far-from-equilibrium processes. 
The exact theory shows that no matter how slowly the symmetry breaking process is driven, defect formation prevents an adiabatic quantum-to-classical time evolution. In a closed system, therefore, a stable symmetry-broken state cannot evolve out of  a symmetric quantum state -- neither spontaneously nor by driving it.

\end{document}